\documentclass[preprint,showpacs,amsmath,amssymb,superscriptaddress,citesort]{revtex4}
\usepackage{graphicx}
\usepackage{dcolumn}
\usepackage{bm}

\newcommand{\nc}{\newcommand}
\nc{\rnc}{\renewcommand}
\nc{\nn}{\nonumber}
\nc{\ch}{\cosh}
\nc{\sh}{\sinh}
\rnc{\th}{\tanh}
\nc{\db}{\displaybreak[0]\\}
\nc{\bra}{\langle}
\nc{\ket}{\rangle}
\rnc{\(}{\left(}
\rnc{\)}{\right)}
\nc{\xxx}{$XXX \,$}
\nc{\lam}{\lambda}
\begin{document}
%
\title{Next Nearest-Neighbor Correlation Functions of the Spin-1/2 XXZ 
Chain at Critical Region}
%
\author{Go Kato}
\email[]{kato@monet.phys.s.u-tokyo.ac.jp}
\affiliation{Department of Physics, Graduate School of Science, University of Tokyo, Hongo 7-3-1, Bunkyo-ku, Tokyo 113-0033, Japan}
\author{Masahiro Shiroishi}
\email[]{siroisi@issp.u-tokyo.ac.jp}
\affiliation{Institute for Solid State Physics, University of Tokyo, Kashiwa,
Chiba 277-8571, Japan}
\author{Minoru Takahashi}
\email[]{mtaka@issp.u-tokyo.ac.jp}
\affiliation{Institute for Solid State Physics, University of Tokyo, Kashiwa,
Chiba 277-8571, Japan}
\author{Kazumitsu Sakai}
\email[]{kaz@issp.u-tokyo.ac.jp}
\affiliation{Institute for Solid State Physics, University of Tokyo, Kashiwa,
Chiba 277-8571, Japan}
\date{April 22, 2003}
%
%
\begin{abstract}
The correlation functions of the spin-${\frac{1}{2}}$ $XXZ$ spin chain 
in the ground state are expressed in the form of the multiple integrals.
For ${-1< \Delta <1}$, they were obtained by Jimbo and Miwa in 1996. 
Especially 
the next nearest-neighbour correlation functions are given as certain 
three-dimensional integrals. We shall show these integrals can be reduced 
to one-dimensional ones and thereby evaluate the values of the 
next nearest-neighbor correlation functions. We have also found that 
the remaining one-dimensinal integrals can be evaluated analytically, 
when ${\nu = \cos^{-1}(\Delta)/\pi}$ is a rational number. 
\end{abstract}

\maketitle
%
%
The spin-1/2 ${XXZ}$ chain is one of the fundamental models in the study 
of the low-dimensional magnetism. The Hamiltonian is given by
\begin{align}
H=\sum_{j=-\infty}^{\infty} 
\left\{ S^x_j S^x_{j+1} + S^y_j S^y_{j+1} 
+ \Delta S^z_j S^z_{j+1} \right\}, \label{XXZ}
\end{align}
where  $S= \sigma /2$ and  $\sigma$ are Pauli matrices.
The model can be solved by Bethe ansatz method and diverse physical properties 
have been investigated with varying the anisotropy parameter $\Delta$ 
\cite{Bethe31,Hulthen38,TakahashiBook,KorepinBook}. Especially, in the region ${-1 < \Delta \le 1}$, the ground state is critically disordered and 
the excitation spectrum is gapless. Then the long-distance asymptotics of two-point the correlation functions such as $\langle S^{\alpha}_j S^{\alpha}_{j+k} \rangle_{k \gg 1} 
\ \ {\alpha=x,y,z}$ 
are shown 
to decay as a power law via field theoretical approach (see, for example, \cite{Lukyanov03} and the references therein). However, if possible, 
it is more desirable  \\ 
\noindent 
(1) to calculate $\langle S^{\alpha}_j S^{\alpha}_{j+k} \rangle$ for finite ${k}$ 
first and \\
\noindent
(2) to derive its asymptotic behavior exactly. \\
Unfortunately such a program has not been achieved except for the ${\Delta=0}$ case \cite{Lieb61,McCoy68}.

In 1996, Jimbo and Miwa \cite{Jimbo96} obtained the multiple integral representation for the arbitrary correlation functions of the ${XXZ}$ chain for ${-1 < \Delta < 1}$. For example, the Emptiness Formation Probability (EFP) 
\cite{Korepin94} defined by 
\begin{align}
P(n) = \Big\langle \prod_{j=1}^n \left(S_j^z + \frac{1}{2} \right)
\Big\rangle, 
\label{EFP}
\end{align}
has the integral representation 
\begin{align}
P(n)&= \left(- \nu \right)^{-\frac{n(n-1)}{2}} \int_{-\infty}^{\infty} 
\frac{{\rm d} x_1}{2 \pi} \cdots \int_{-\infty}^{\infty} 
\frac{{\rm d} x_n}{2 \pi}   \prod_{a>b} \frac{\sinh (x_a-x_b)}
{\sinh \left( \left(x_a-x_b - {\rm i} \pi \right) \nu \right)}\nonumber \\
&  \ \ \ \ \ \ \ \ \ \ \ \ \ \ \ \ \ \ \ \ \times  \prod_{k=1}^{n} \frac{\sinh^{n-k} \left( \left(x_k + {\rm i} \pi/2 \right) \nu \right) 
\sinh^{k-1} \left( \left(x_k - {\rm i} \pi/2 \right) \nu \right)}{\cosh^n x_k},
\label{EFPintegral}
\end{align}
where the  parameter ${\nu}$ is related to the anisotropy $\Delta$ as
\begin{align}
\nu = \frac{1}{\pi} \cos^{-1}(\Delta). \label{nu}
\end{align} 
Recently there have been an increasing number of researches concerning the properties of the EFP \cite{Korepin94,Essler95n1,Essler95n2,Kitanine00,Razumov01,Kitanine02n1,Shiroishi01,Boos01,Boos02,BKNS02,Boos03,Abanov02,Korepin03,Kitanine02n2}. Particularly, in the isotropic limit ${\Delta \to 1 (\nu \to 0)}$, the general method to evaluate the multiple integral was recently developed by Boos and Korepin \cite{Boos01,Boos02,BKNS02}. 

It is of some importance to note that ${P(2)}$ and ${P(3)}$ are related to the nearest and next nearest-neighbor correlation functions :
\begin{align}
\langle S_{j}^z S_{j+1}^z \rangle &= P(2)-1/4, \label{P2} \\ 
\langle S_{j}^z S_{j+2}^z \rangle &= 2 (P(3)-P(2)+1/8). \label{P3} 
\end{align}  
One of our purposes in this letter is to evaluate ${\langle S_{j}^z S_{j+2}^z \rangle}$ through the integral representations of ${P(3)}$ and ${P(2)}$. 
Similarly the nearest and the next nearest-neighbor transverse correlation functions have the integral representations,
\begin{align}
\langle S_{j}^x S_{j+1}^x \rangle = -\frac{1}{2 \nu}  \int_{-\infty}^{\infty}  \frac{{\rm d} x_1}{2 \pi} \int_{-\infty}^{\infty} \frac{{\rm d} x_2}{2 \pi} 
\frac{\sinh (x_2-x_1)}{\sinh \left( \left(x_2-x_1 \right) \nu \right)} 
\frac{\sinh \left( \left( x_1 + {\rm i} \pi/2 
\right) \nu \right) \sinh \left( \left(x_2-{\rm i} \pi/2 \right) 
\nu \right)}{\cosh^2 x_1 \cosh^2 x_2}, 
\label{transverse1}
\end{align}
and 
\begin{align}
\langle S_{j}^x S_{j+2}^x \rangle &= - \frac{1}{\nu^3} 
\prod_{k=1}^{3} \int_{-\infty}^{\infty} \frac{{\rm d} x_k}{2 \pi} 
\frac{\sinh^{3-k} \left( \left(x_k + {\rm i} \pi/2 
\right) \nu \right) \sinh^{k-1} \left( \left(x_k - {\rm i} \pi/2 \right) 
\nu \right)}{\cosh^3 x_k} \nonumber \\
& \hspace{2cm} \times  \frac{\sinh (x_2-x_1)}
{\sinh \left(\left(x_2-x_1 \right) \nu \right)}  
\frac{\sinh (x_3-x_1)}
{\sinh \left(\left(x_3-x_1 \right) \nu \right)}  
\frac{\sinh (x_3-x_2)}
{\sinh \left( \left(x_3-x_2 - {\rm i} \pi \right) \nu \right)}, 
\label{transverse2}
\end{align}
respectively.

It was already shown by Jimbo and Miwa that the two-dimensional integral 
for ${P(2)}$ reduces to the one-dimensional one for arbitrary ${\nu}$ as
\begin{align}
& P(2)= \frac{1}{2} + \frac{1}{2 \pi^2 \sin \pi \nu} \frac{\partial}{\partial \nu} 
\left\{ \sin \pi \nu \int_{-\infty}^{\infty} \frac{\sinh(1-\nu) w}{\sinh w \cosh \nu w} 
{\rm d} w \right\}, \label{P2new}
\end{align}
which leads to the one-dimensional integral representation of  
${\langle S_{j}^z S_{j+1}^z \rangle}$ via the relation (\ref{P2}). 
The result coincides with a different derivation of ${\langle S_{j}^z S_{j+1}^z \rangle}$ from the ground-state energy per site ${e_0}$ : 
\begin{align}
\langle S_{j}^z S_{j+1}^z \rangle &=  \frac{\partial e_0}{\partial \Delta} 
= - \frac{1}{\pi \sin \pi \nu} \frac{\partial e_0}{\partial \nu} \nonumber \\
&= \frac{1}{4} + \frac{\cot \pi \nu}{2 \pi} 
\int_{-\infty}^{\infty} \frac{{\rm d} w }{\sinh w} \frac{\sinh(1-\nu) w}{ \cosh \nu w} - \frac{1}{2 \pi^2} 
\int_{-\infty}^{\infty} \frac{{\rm d} w}{\sinh w} \frac{ w \cosh w}{(\cosh \nu w)^2}, \label{nextZZ}
\end{align}
where 
\begin{align}
e_0 =\frac{\Delta}{4} - \frac{\sin \pi \nu}{2 \pi} \int_{-\infty}^{\infty} 
\frac{\sinh(1- \nu) w}{\sinh w \cosh \nu w} {\rm d} w. 
\end{align}
Similarly we can get the one-dimensional integral representation for 
${\langle S_{j}^x S_{j+1}^x \rangle }$ as 
\begin{align}
\langle S_{j}^x S_{j+1}^x \rangle 
&= \frac{1}{2} 
\left( e_0 - \Delta \langle S_{j}^z S_{j+1}^z \rangle \right)  
\nonumber \\
&= - \frac{1}{4 \pi \sin \pi \nu} 
\int_{-\infty}^{\infty} \frac{{\rm d} w }{\sinh w} \frac{\sinh(1-\nu) w}{ \cosh \nu w}   + \frac{\cos \pi \nu}{4 \pi^2} 
\int_{-\infty}^{\infty} \frac{{\rm d} w}{\sinh w} \frac{ w \cosh w}{(\cosh \nu w)^2}.  \label{nextXX}
\end{align}
Thus, for the nearest-neighbor correlation functions, we have known 
the two-dimensional integrals by Jimbo-Miwa formula can be reduced to 
one-dimensional ones. The main result of this letter is that 
{\it three-dimensional integrals for ${P(3)}$ and 
${\langle S_{j}^x S_{j+2}^x \rangle}$ can also be reduced 
to one-dimensional integrals}. In other words, we have succeeded in performing 
the integrals for ${P(3)}$ and ${\langle S_{j}^x S_{j+2}^x \rangle}$  twice. 
Our results are 
\begin{align}
P(3)&= \frac{1}{2} 
+ \int_{- \infty- {\rm i} \delta }^{\infty- {\rm i} \delta} 
\frac{{\rm d} x}{\sinh x} \Bigg[ \frac{1- \cos 2 \pi \nu}{16 \pi^2} 
\frac{\partial}{\partial \nu} 
\left\{ \frac{\cosh 3 \nu x}{(\sinh \nu x)^3} \right\} 
+ \frac{3 \tan \pi \nu}{8 \pi} \frac{\cosh 3\nu x}
{(\sinh \nu x)^3} \nonumber \\
&  \hspace{3cm} - \frac{4- \cos 2 \pi \nu}{4 \pi^2} 
\frac{\partial \left(\coth \nu x \right)}{\partial \nu} 
- \frac{\left(4- \cos 2 \pi \nu \right) \coth \nu x}
{2 \pi \sin 2 \pi \nu} \Bigg], \label{P3new}
\end{align}
and 
\begin{align}
\langle S_{j}^x S_{j+2}^x \rangle 
&= \int_{-\infty- {\rm i} \delta}^{\infty - {\rm i} \delta} 
\frac{{\rm d} x}{\sinh x} \Bigg[ 
- \frac{\left(\sin \pi \nu \right)^2}{8 \pi^2} 
\frac{\partial}{\partial \nu} 
\left\{ \frac{\cosh 3 \nu x}{(\sinh \nu x)^3} \right\} 
- \frac{3 \cos 2 \pi \nu(1- \cos 2 \pi\nu)}{8 \pi \sin 2 \pi \nu} 
\frac{\cosh 3 \nu x}{(\sinh \nu x)^3} \nonumber \\ 
& \hspace{1.5cm} + \frac{1}{4 \pi^2} 
\frac{\partial \left(\coth \nu x \right)}{\partial \nu} 
+ \frac{\left\{ 1+ 3\cos 2 \pi \nu -3 \left(\cos 2 \pi \nu \right)^2 
\right\} \coth \nu x}
{2 \pi \sin 2 \pi \nu} 
\Bigg]. \label{transnew}
\end{align}
Here ${\delta \ (0<|\delta| <\pi)}$ should take some non-zero 
real value, which is introduced to avoid the singularity at 
the origin. We, however, note that the singular term may be 
subtracted from the integrand in principle, as its residue 
vanishes.  Or alternatively, by applying the Fourier 
transform, we can express the one-dimensional representations 
(\ref{P3new}) and (\ref{transnew}) as 
\begin{align}
P(3)&= \frac{1}{2} + \int_{- \infty}^{\infty} 
\frac{{\rm d} w}{\sinh w} \frac{\sinh (1-\nu)w}{\cosh \nu w} 
\left[ \frac{1+ 2 \cos 2 \pi \nu}{2 \pi \sin 2 \pi \nu}
+ \frac{3 \tan \pi \nu}{4 \pi^3} w^2 \right] \nonumber \\
& \ \ \  - \int_{- \infty}^{\infty} 
\frac{{\rm d} w}{\sinh w} \frac{\cosh w}{(\cosh \nu w)^2} 
\left[ \frac{3}{4 \pi^2} w + \frac{(\sin \pi \nu)^2}{4 \pi^4} 
w^3 \right], \label{P3w}
\end{align}
and
\begin{align}
\langle S_{j}^x S_{j+2}^x \rangle &= -\int_{- \infty}^{\infty} 
\frac{{\rm d} w}{\sinh w} \frac{\sinh (1-\nu)w}{\cosh \nu w} 
\left[ \frac{1}{2 \pi \sin 2 \pi \nu}
+ \frac{3 \cos 2 \pi \nu \tan \pi \nu}{4 \pi^3} w^2 \right] \nonumber \\
& \ \ \  + \int_{- \infty}^{\infty} 
\frac{{\rm d} w}{\sinh w} \frac{\cosh w}{(\cosh \nu w)^2} 
\left[ \frac{\cos 2 \pi \nu}{4 \pi^2} w + \frac{(\sin \pi \nu)^2}{4 \pi^4} 
w^3 \right]. \label{transw}
\end{align}
Note that in the representations (\ref{P3w}) and (\ref{transw}), there are 
no singularities at the origin. This is in the similar situation as (\ref{nextZZ}) and (\ref{nextXX}). Combining (\ref{P2new}) and (\ref{P3w}), we can also 
write down the one-dimensional integral representation for ${\langle S_{j}^z 
S_{j+2}^z \rangle}$ through the relation (\ref{P3}),
\begin{align}
\langle S_{j}^z S_{j+2}^z \rangle &= \frac{1}{4}+ \int_{- \infty}^{\infty} 
\frac{{\rm d} w}{\sinh w} \frac{\sinh (1-\nu)w}{\cosh \nu w} 
\left[ \frac{\cot 2 \pi \nu}{\pi}
+ \frac{3 \tan \pi \nu}{2 \pi^3} w^2 \right] \nonumber \\
& \ \ \  - \int_{- \infty}^{\infty} 
\frac{{\rm d} w}{\sinh w} \frac{\cosh w}{(\cosh \nu w)^2} 
\left[ \frac{1}{2 \pi^2} w + \frac{(\sin \pi \nu)^2}{2 \pi^4} 
w^3 \right]. \label{longiw}
\end{align}

Now let us discuss some properties of the obtained one-dimensional representations (\ref{transw}) and (\ref{longiw}). 
At a first glance, some integrands in (\ref{transw}) and (\ref{longiw}) are divergent at ${\nu=1/2}$ and also in the limit ${\nu \to 0}$ and ${\nu \to 1}$, due to the the factor ${1/\sin{2 \pi \nu}}$. However, by investigating the integrands carefully, we have found these singular terms cancel each other, therefore yielding  definite finite values for the correlation functions : 
\begin{itemize}
\item ${\nu=1/2}$ : 
\begin{align}
\langle S_{j}^x S_{j+2}^x \rangle = \dfrac{1}{\pi^2}, \ \ \ \ 
\langle S_{j}^z S_{j+2}^z \rangle = 0, \label{Delta0}
\end{align}

\item ${\nu \to 0}$ :  
\begin{align}
\langle S_{j}^x S_{j+2}^x \rangle, \ \ 
\langle S_{j}^z S_{j+2}^z \rangle \to \frac{1}{12}- \frac{4}{3} \ln 2
+ \frac{3}{4} \zeta(3), 
 \label{Delta1}
\end{align}

\item ${\nu \to 1}$ : 
\begin{align}
\langle S_{j}^x S_{j+2}^x \rangle \to \dfrac{1}{8}, \ \ 
\langle S_{j}^z S_{j+2}^z \rangle \to 0. \label{Deltam1}
\end{align}
\end{itemize}
 It is especially intriguing to observe that the ${\nu \to 0}$ limit 
(\ref{Delta1}) reproduces the known result by one of the authors 
in \cite{Takahashi77}. 

\begin{figure}[htbp]
  \includegraphics[width=10cm]{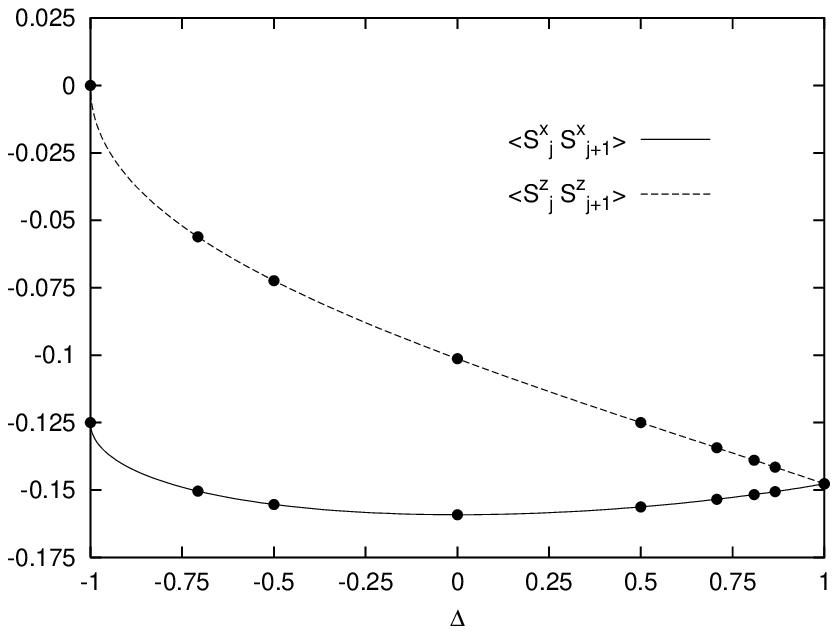}
\caption{\label{fig:1} Nearest-neighbor correlation functions for the XXZ chain}
\end{figure}

\begin{figure}[htbp]
  \includegraphics[width=10cm]{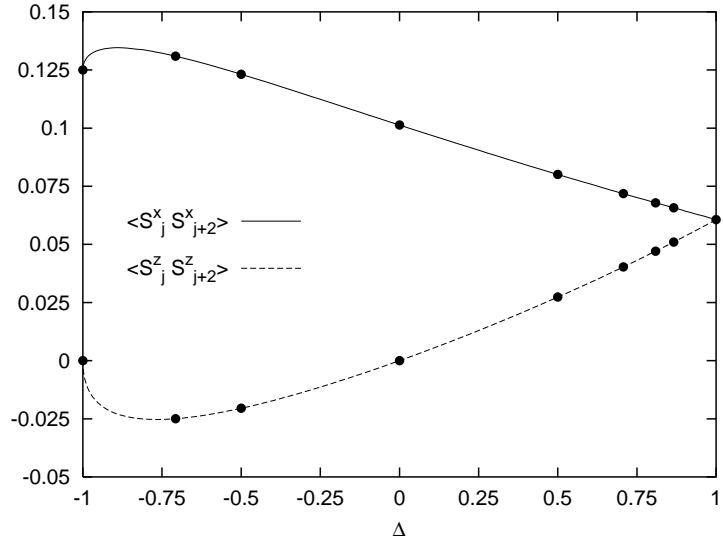}
\caption{\label{fig:2} Next nearest-neighbor correlation functions for the XXZ chain}
\end{figure}

More generally, we have found when ${\nu}$ takes a rational value, the one-dimensional integrals can be evaluated analytically. Some of our explicit results are summarized in Table 1. One can see that the correlation functions are in general a polynomial of ${1/\pi}$. Particularly, when ${\nu=1/3}$, we expect all the correlation functions are given solely as a single rational number 
(c.f. \cite{Razumov01,Kitanine02n1}). In Fig.~\ref{fig:1} and Fig.~\ref{fig:2}, plotted are the numerical values of the nearest and the next nearest-neighbor correlation functions calculated from the one-dimensional integral representations. For comparison, the analytical values in Table 1 are represented by the filled circles. 

We have obtained (\ref{P3new}) and (\ref{transnew}) by generalizing the method 
developed by Boos and Korepin \cite{Boos01,Boos02}, which allows us to 
calculate the multiple integrals for the correlation functions of the 
${XXX}$ model (${\Delta=1}$). Below we outline  the derivation of (\ref{P3new}) and (\ref{transnew}) quite briefly. The details of the calculations will be published in a separate paper \cite{Kato03}. 
First we introduce the following convenient notation for ${P(3)}$ :
\begin{align}
P(3) =  \prod_{k=1}^{3} 
\int_{-\infty - {\rm i}/2}^{\infty-{\rm i}/2} \frac{{\rm d} \lambda_{k}}{2 \pi {\rm i}} 
U_3(\lambda_1,\lambda_2,\lambda_3) T_3(\lambda_1,\lambda_2,\lambda_3),
\label{notation} 
\end{align}
where
\begin{align}
U_3(\lambda_1,\lambda_2,\lambda_3) =
\frac{\pi^3 \prod_{1 \le k< j \le3} \sinh \pi \left(\lambda_j-\lambda_k \right)}
{\nu^3 \prod_{j=1}^{3} \sinh^3 \pi \lambda_j}, 
\end{align}
and 
\begin{align}
T_3(\lambda_1,\lambda_2,\lambda_3) =
\frac{(q z_1-1)^2 (q z_2-1)(z_2-1) (z_3-1)^2}
{8 (z_2-q z_1)(z_3-q z_1)(z_3-q z_2)}, 
\end{align}
with ${q   \equiv  {\rm e}^{2 \pi {\rm i} \nu},
z_i  \equiv  {\rm e}^{2 \pi \nu {\lambda_i}}, \ (i=1,2,3)}$.
Then after similar but more complicated  calculation as Boos and Korepin 
\cite{Boos02}, we can transform the integrand 
${ T_3(\lambda_1,\lambda_2,\lambda_3)}$ into a certain 
{\it canonical form} without changing the value of the integral as
\begin{align}
T_c = P_0 + \frac{P_1}{z_2-z_1}, 
\end{align}
 where
\begin{align}
P_0 =& \frac{(1+q)^2}{8q}\frac{z_1}{z_3}, \\
P_1 =& \frac{3(1+q)}{2 q} - \frac{(1+10q+q^2)z_1}{8 q}
- \frac{3(1+q)^2}{8q^2 z_1}  \nonumber \\
&+ z_3 \left\{ - \frac{1+10q+q^2}{8q} + \frac{3(1+q) z_1}{8}+ 
\frac{3(1+q)}{8q z_1} \right\} \nonumber \\
&+ \frac{1}{z_3} \left\{ - \frac{3(1+q)^2}{8q^2} + \frac{3(1+q) z_1}{8q}+ 
\frac{(1+q)(1+q+q^2)}{8q^3 z_1} \right\}. 
\end{align}
We find the first part of ${T_c}$, namely ${P_0}$, can be integrated 
easily 
\begin{align}
\prod_{k=1}^{3}  \int_{-\infty - {\rm i}/2}^{\infty-{\rm i}/2} 
\frac{{\rm d} \lambda_{k}}{2 \pi {\rm i}} U_3(\lambda_1,\lambda_2,\lambda_3) 
\frac{(1+q)^2 {\rm e}^{2 \pi \nu (\lambda_1-\lambda_3)}}{8q}
= \frac{1}{2}. \label{P0integral}
\end{align}
The second part ${P_1/(z_2-z_1)}$ can be integrated twice, 
namely with respect to ${\lambda_3}$ and ${ a \equiv (\lambda_2 
+ \lambda_1)/2 }$. 
Finally the remaining one-dimensional integration with respect to 
${ d \equiv (\lambda_2-\lambda_1)/2}$  together with (\ref{P0integral}) 
gives the expression (\ref{P3new}). 

Similarly for ${\langle S_{j}^x S_{j+2}^x \rangle}$, 
${T_3(\lambda_1,\lambda_2,\lambda_3)}$ in (\ref{notation}) 
is replaced by 
\begin{align}
T_3(\lambda_1,\lambda_2,\lambda_3) =
\frac{(q z_1-1)^2 (q z_2-1)(z_2-1) (z_3-1)^2}
{8q (z_2-z_1)(z_3-z_1)(z_3-q z_2)}. 
\end{align}
As a corresponding canonical form, we have found 
\begin{align}
T_c = \frac{P_1}{z_2-z_1}, 
\end{align}
with
\begin{align}
P_1 =& - \frac{(1+q)(3+q^2)}{8q^2} + \frac{(3- 2q+3 q^2) z_1}{8q} 
+ \frac{(1+q)^2}{8 q^3 z_1} \nonumber \\
&+ z_3 \left\{ \frac{3-q}{4q} - \frac{(1+q) z_1}{8q}+ 
\frac{(1+q)(-2+q)}{8q^2 z_1} \right\} \nonumber \\
&+ \frac{1}{z_3} \left\{\frac{1+q}{2q} - \frac{(1+q) z_1}{8}- 
\frac{1+q}{8q^2 z_1} \right\}. 
\end{align}
Again after integrating with respect to ${\lambda_3}$ and 
${ a \equiv (\lambda_2 + \lambda_1)/2 }$, we arrive at 
the one-dimensional integral representation (\ref{transnew}).

In conclusion, we have shown the multiple integrals for the correlation 
functions of the ${XXZ}$ chain at the critical region ${-1<\Delta<1}$,  
can be reduced to the one-dimensional ones in the case of the next 
nearest-neighbor correlation functions. This property will be generalized 
to other higher-neighbor correlations as well as the correlations 
in the massive region (${\Delta>1}$) \cite{Jimbo92,JimboBook}. 
In this respect, we like to refer to the recent work by Boos, Korepin 
and Smirnov \cite{BKS03}, where they have shown the reducibility of the multiple integrals in the case of the ${XXX}$ chain. 

In our future work, we are particularly interested in calculating the 
third-neighbor correlation functions ${\langle S_{j}^x S_{j+3}^x \rangle}$ 
and ${\langle S_{j}^z S_{j+3}^z \rangle}$ for general ${\Delta}$. For ${\Delta=1}$, they were recently calculated in \cite{Sakai03} from the multiple integrals  \cite{Korepin94,Nakayashiki94} by Boos-Korepin method .

\vspace{12pt}
We are grateful to H.E.~Boos, M.~Jimbo, V.E.~Korepin, Y.~Nishiyama, J.~Suzuki and M.~Wadati for valuable discussions. This work is in part supported by Grant-in-Aid for the Scientific Research (B) No.~14340099 from the Ministry of Education, Culture, Sports, Science and Technology, Japan. GK and KS are supported by the JSPS research 
fellowships for young scientists. MS is supported by Grant-in-Aid for young scientists No.~14740228.

\begin{table}
\begin{center}
\begin{tabular}{|c||c|c|c|c|}
\hline
${\displaystyle \nu}$ & 
$\langle S_j^xS_{j+1}^x \rangle$
&
$\langle S_j^z S_{j+1}^z \rangle$
&
$\langle S_j^xS_{j+2}^x \rangle$
& 
$\langle S_j^zS_{j+2}^z \rangle$\\
\hline
\scriptsize{$0$} &
$\frac{1}{12}-\frac{\ln 2}{3}$&$\frac{1}{12}-\frac{\ln 2}{3}$&
$\frac{1}{12}-\frac{4\ln2}{3}+ \frac{3\zeta(3)}{4}$&
$\frac{1}{12}-\frac{4\ln2}{3}+ \frac{3\zeta(3)}{4}$\\
\hline 
$\frac{1}{2}$&
$-\frac{1}{2 \pi}$&$-\frac{1}{\pi^2}$&$\frac{1}{\pi^2}$
&\scriptsize{$0$} \\
\hline
$\frac{1}{3}$& $-\frac{5}{32}$&$-\frac{1}{8}$&$\frac{41}{512}$
&$\frac{7}{256}$\\
\hline
$\frac{1}{4}$
&$-\frac{\sqrt{2}}{2 \pi}+
\frac{\sqrt{2}}{2 \pi^2}$
&
$-\frac{1}{4}+\frac{1}{\pi}-\frac{2}{\pi^2}$
&$\frac{3}{16} - \frac{1}{\pi} + \frac{2}{\pi^2}$&
$-\frac{5}{8}+\frac{4}{\pi}-\frac{6}{\pi^2}$\\
\hline
$\frac{1}{5}$&$-\frac{3}{64} - \frac{3 \sqrt{5}}{64}$
&$-\frac{19}{8}$ +\scriptsize{$\sqrt{5}$}
&$\frac{7737}{1024} - \frac{3429 \sqrt{5}}{1024}$
&$-\frac{3529}{512} + \frac{1589 \sqrt{5}}{512}$\\
\hline
$\frac{1}{6}$
&$\frac{\sqrt{3}}{48} - \frac{1}{\pi} + 
\frac{3 \sqrt{3}}{4 \pi^2}$
&$-\frac{7}{18} + \frac{\sqrt{3}}{\pi}-\frac{3}{\pi^2}$
&$\frac{247}{576} 
- \frac{17 \sqrt{3}}{12 \pi} +\frac{33}{ 8 \pi^2}$
&$-\frac{283}{288}+\frac{11 \sqrt{3}}{3 \pi}
- \frac{39}{4 \pi^2}$
\\
\hline
$\frac{2}{3}$&$\frac{47}{128}-\frac{\sqrt{3}}{4}-\frac{9}{32 \pi}$
&
$\frac{23}{32} 
- \frac{\sqrt{3}}{4} - \frac{9}{8 \pi}$
&$ - \frac{2719}{8192} +\frac{47 \sqrt{3}}{256} + \frac{441}{1024 \pi}$
& $\frac{8671}{4096} 
- \frac{49 \sqrt{3}}{64} - \frac{1305}{512 \pi}$\\
\hline
$\frac{3}{4}$&
\begin{tabular}{l}
$- \frac{4 \sqrt{6}}{27}
+ \frac{64 \sqrt{2}}{243} - \frac{\sqrt{2}}{6\pi}$ \\ 
$\qquad-\frac{8 \sqrt{6}}{81\pi} - \frac{\sqrt{2}}{18 \pi^2} $
\end{tabular}&
\begin{tabular}{l}
$\frac{781}{972} 
-\frac{8 \sqrt{3}}{27} -\frac{1}{3\pi}$ \\ 
$\quad-\frac{32 \sqrt{3}}{81\pi}-\frac{2}{9 \pi^2}$
\end{tabular}&
\begin{tabular}{l}
$-\frac{22111}{34992}
+ \frac{8 \sqrt{3}}{27} + \frac{1}{3\pi}$ \\ 
$\qquad+ \frac{160 \sqrt{3}}{729\pi} + \frac{2}{9 \pi^2}$
\end{tabular}&
\begin{tabular}{l}
$\frac{36169}{17496}
-\frac{160 \sqrt{3}}{243} -\frac{4}{3\pi}$\\ 
$\qquad-\frac{608 \sqrt{3}}{729\pi}- \frac{2}{3 \pi^2}$
\end{tabular}
\\
\hline
\scriptsize{1}&$-\frac{1}{8}$&\scriptsize{0}&$ \frac{1}{8}$
&\scriptsize{0}
\\
\hline
\end{tabular}
\end{center}
\caption{Some analytical values of the correlation functions when ${\nu}$ takes   rational values }
\label{table1}
\end{table}

%


\begin{thebibliography}{}

\bibitem{Bethe31} H.A.~Bethe,
Z. Phys. {\bf 71} (1931) 205.

\bibitem{Hulthen38} L.~Hulth\'{e}n,
Ark. Mat. Astron. Fys. A {\bf 26} (1938) 1.

\bibitem{TakahashiBook} M.~Takahashi, 
\textit{Thermodynamics of One-Dimensional Solvable Models}, 
(Cambridge University Press,Cambridge, 1999).

\bibitem{KorepinBook} V.E~Korepin, N.M.~Bogoliubov and A.G.~Izergin,
\textit{Quantum Inverse Scattering Method and Correlation Functions},
(Cambridge University Press, Cambridge, 1993).

\bibitem{Lukyanov03} S.~Lukyanov and V.~Terras,
Nucl. Phys. B {\bf 654} (2003) 323.

\bibitem{Lieb61} E.H.~Lieb, T.~Schultz and D.~Mattis,
A. Phys. (N.Y.) {\bf 16} (1961) 417. 

\bibitem{McCoy68} B.M.~McCoy,
Phys. Rev. {\bf 173} (1968) 531.

\bibitem{Jimbo96} M.~Jimbo and T.~Miwa, 
J. Phys. A {\bf 29} (1996) 2923.

\bibitem{Korepin94} V.E.~Korepin, A.G.~Izergin, F.H.L.~Essler and D.B.~Uglov
Phys. Lett. A {\bf 190} (1994) 182.

\bibitem{Essler95n1} F.H.L.~Essler, H.~Frahm, A.G.~Izergin and V.E.~Korepin, 
Commun. Math. Phys. {\bf 174} (1995) 191.

\bibitem{Essler95n2} F.H.L.~Essler, H.~Frahm,  A.R.~Its and V.E.~Korepin, 
Nucl Phys {\bf B 446}, (1995) 448.

\bibitem{Kitanine00} N.~Kitanine, J. M.~Maillet and V.~Terras,
Nucl. Phys. {\bf B 567} (2000) 554.

\bibitem{Razumov01} A.V.~Razumov and Yu.G.~Stroganov, 
J. Phys. A: Math. Gen. {\bf 34} (2001) 3185, \textit{ibid.} 5335.

\bibitem{Kitanine02n1} N.~Kitanine, J.M.~Maillet, N.A.~Slavnov,
and V.~Terras, 
J. Phys. A. {\bf 35} (2002) L385.

\bibitem{Shiroishi01} M.~Shiroishi, M.~Takahashi and Y.~Nishiyama,
J. Phys. Soc. Jpn. {\bf 70} (2001) 3535.

\bibitem{Boos01} H.E.~Boos and V.E.~Korepin,
J. Phys. A {\bf 34} (2001) 5311.

\bibitem{Boos02} H.E.~Boos and V.E.~Korepin,
\textit{Integrable models and Beyond}, edited by 
 M.~Kashiwara and T.~Miwa (Birkh\"auser, Boston, 2002); hep-th/0105144.

\bibitem{BKNS02} H.E.~Boos, V.E.~Korepin, Y.~Nishiyama and M.~Shiroishi,
J. Phys. A {\bf 35} (2002) 4443.

\bibitem{Boos03} H.E.~Boos, V.E.~Korepin and F.A.~Smirnov,
Nucl. Phys. B. {\bf 658} (2003) 417.

\bibitem{Abanov02} A.G.~Abanov and V.E.~Korepin,
Nucl. Phys. {\bf B 647} (2002) 565.

\bibitem{Korepin03}  V.E.~Korepin, S.~Lukyanov, Y.~Nishiyama and 
M.~Shiroishi, hep-th/0210140, to appear in Phys. Lett. A (2003).

\bibitem{Kitanine02n2} N.~Kitanine, J.M.~Maillet, N.A.~Slavnov,
and V.~Terras, J. Phys. A. {\bf 35} (2002) L753.

\bibitem{Takahashi77}  M.~Takahashi, J. Phys. C {\bf 10} (1977) 1289.

\bibitem{Kato03} 
G.~Kato, K.~Sakai, M.~Shiroishi and M.~Takahashi,
in preparation.

\bibitem{Jimbo92} M.~Jimbo, K.~Miki, T.~Miwa and A.~Nakayashiki,
Phys. Lett. A {\bf 168}, (1992) 256.

\bibitem{JimboBook} M.~Jimbo and T.~Miwa,
\textit{Algebraic Analysis of
Solvable Lattice Models}, (American Mathematical Society,
Providence, RI, 1995). 

\bibitem{BKS03}
H.~Boos, V.E.~Korepin and F.~Smirnov,
\textit{New Formulae for the solutions of quantum Knizhnik-Zamolodchikov 
equations on level-4}, hep-th/0304077.

\bibitem{Sakai03}
K.~Sakai, M.~Shiroishi, Y.~Nishiyama and M.~Takahashi,
\textit{Third Neighbor Correlators of one-dimensional
Spin-1/2 Heisenberg Antiferromagnet}, cond-mat/0302564.

\bibitem{Nakayashiki94} 
A.~Nakayashiki,
Int. J. Mod. Phys. A {\bf 9}, (1994) 5673.

\end{thebibliography}
\end{document}